\documentclass[doublespace,aip,jcp,preprint]{revtex4-1}
\usepackage{pslatex}
\usepackage[dvips]{epsfig}
\usepackage[dvips]{graphicx}
\usepackage{subfigure}
\usepackage{multirow}
\usepackage{chemarr}
\usepackage{amssymb}
\usepackage{color}
\begin{document}

\title{Exploring the properties of valence electron based potential functions for the nonbonded 
interactions in atomistic force fields}

\author{Nuria Plattner}
\affiliation{Department of Mathematics and Computer Science, Free University Berlin, 
Arnimallee 6, 14195 Berlin, Germany}
 \email{nuria.plattner@fu-berlin.de}

\date{\today}

\begin{abstract}
\noindent  The possibility to construct and parametrize the nonbonded interactions in atomistic force fields based on the valence electron structure of molecules is explored in this paper. Three different charge distribution models using simple valence electron based potential functions are introduced and compared. It is shown that the three models can be 
constructed such that they only require one adjustable parameter for the electrostatic potential of a molecule. The accuracy of the electrostatic potential is evaluated for the three models and compared to population-derived charges and higher order multipole moments for a set of 12 small molecules. Furthermore the accuracy and parametrization of the interaction energies of the three models is evaluated based on {\it ab initio} intermolecular interaction energies. It is shown that the valence electron potential models provide systematic advantages 
over conventional point charge models for the calculation of intermolecular interaction energies even with the very simple
potential functions used here.
\end{abstract}

\maketitle


\section{Introduction}

\noindent Empirical atomistic force fields\cite{charmm22,amber,GROMOS,OPLS} are a well established tool to calculate interatomic
interactions in atomistic simulations.\cite{Karplus90,vanGunstern94} Despite the high popularity of empirical 
 force fields, commonly used force field functions are still not sufficiently accurate for many applications. 
In order to improve the force field accuracy, various methods have been proposed based on adding new terms to existing 
force field potential functions. The proceeding of adding terms without modifying existing terms implies that existing force field functions 
are an optimal or at least necessary basis to describe interatomic interactions. This assumption is however questionable 
since the interaction potential used in common empirical force fields 
is not systematically derived from the quantum mechanical description of interatomic interactions, but 
determined based on the requirement to reproduce a number of molecular properties, while 
being simple and robust to allow fast calculations of large systems. A popular strategy is 
to divide the interaction potential $V_{ff}$ into bonded and nonbonded terms, where the bonded terms include bonds, 
angles and dihedrals, and the nonbonded terms include electrostatics described by point charges and a Lennard-Jones (LJ) 
potential to represent dispersion and Pauli repulsion, as shown in Equation \ref{ff_eq}.

\begin{equation}
V_{ff} = V_{bonds} + V_{angles} + V_{dihedrals} + V_{elstat} + V_{LJ}
\label{ff_eq}
\end{equation}

\noindent While it is relatively straightforward with this function to fit parameters in 
order to obtain correct molecular geometries, obtaining optimal parameters for intermolecular 
interactions is more involved and has been found dissatisfying in many cases. Therefore different methods have been 
proposed to address this shortcoming, including more complex electrostatic interaction potentials using higher order 
multipole moments\cite{Price92,HodgesStone97,Sokalski01,Price06,JMolM09,KramerMeuwly2013,Popelier2014} or additional 
charge sites\cite{TIP5P,StraubKarplus1991,SaxenaSept2013} as well as polarizabilities\cite{LopesMacKerell2013,AMOEBA,SIBFA}. Atomic multipole 
moments and additional charge sites are just two different mathematical 
descriptions of the anisotropic electrostatic potential around atoms. Therefore a given electrostatic potential can 
always be expressed likewise in terms of higher order multipole moments or additional 
charges.\cite{DevereuxMeuwly2014,Stone96} The supplementary interaction sites in force fields with more complex 
electrostatics potentials or polarizabilities have no direct physical meaning, but arise due to the requirement of 
having more accurate interaction potentials. This raises the question whether these  sites could be chosen based on the electronic structure of molecules. 
This would require each site to represent either the potential functions of nuclei or of electrons. 
In atomistic force fields function (Equation \ref{ff_eq}), the electrons are summed into the 
nuclei. The lack of explicit electron potential representations in empirical force fields is motivated
by the fact that a purely classical representation of electrons is not meaningful since they are 
highly delocalized and their dynamics is dominated by quantum effects. Despite these difficulties a 
force field including electrons, the electron force field (eFF), has been derived from first principles and has been shown to provide meaningful results for 
a number of applications.\cite{SuGoddard2007} In eFF, the energy $V_{eFF}$ is the sum of a Hartree 
product kinetic energy ($V_{kin}$), a Hartree product electrostatic energy decomposed into interactions between nuclei ($nuc$) and electrons ($elec$), and an antisymmetrization (Pauli) 
correction: 

\begin{align}
V_{eFF}= V_{kin}+V_{nuc-nuc}+V_{nuc-elec}+V_{elec-elec}+V_{Pauli}
\label{eff_eq}
\end{align}

\noindent The electrons are described by scalable Gaussian wavepackets. The size of
the wavepacket is determined in each step self-consistently by minimizing the total potential and
kinetic energy. eFF is designed for large systems with various excited electrons and works without specific atom type 
parameters. It would however not be suited as a force field for biomolecules in the ground state as its not sufficiently 
accurate and calculations with eFF are substantially more expensive than with conventional force fields. The development 
of eFF demonstrates that a relatively simple force field incorporating the dynamics of nuclei and electrons can be 
derived from first principles. At the same time the shortcomings of eFF 
underline the benefits of building force fields based on empirical parameters. While for a force field derived from first 
principles a high level of complexity, e.g. in the choice of basis set, is required to obtain correct values for simple 
observables such as bond lengths, this is trivially achieved in empirical force fields by using such observables as 
parameters. The price paid for this is the high number of parameters to be determined, resulting in a substantial 
effort in force field parametrization.\\

\noindent Given the up- and downsides of both types of force fields, the question arises whether there are advantages 
of constructing a force field based on concepts similar to eFF, but incorporating simple observables as empirical 
parameters in order to improve the force field accuracy. In the following this possibility will be explored for the 
nonbonded interactions in atomistic force fields. The aim is to assess simple ways of incorporating valence electron based 
potentials into interaction models without aiming at explicitly representing the dynamics of electrons, but rather as a 
means of describing and improving intermolecular interactions. In the following, 
three interaction models based on this idea will be introduced and compared. One of the models contains no additional
interaction sites and can therefore be used for comparison with conventional force fields. For the other two 
models additional interaction sites representing electron potentials are introduced. While in eFF scalable Gaussian 
functions are  used for the representation of electrons, the electron potentials in the three models 
are either described by point charges or by spherical Gaussian functions of a fixed width. Having Gaussian functions of a
fixed width means computationally that a self-consistent variational calculation is not required. 
This approach is meaningful because the electron potentials are only used for the nonbonded interactions here and the bond
length is provided as a parameter. In eFF chemical bonding is the result of competing kinetic and potential energy terms 
of electrons which determine the width of the Gaussian functions self-consistently.\cite{SuGoddard2007} To further 
simplify the models, potentials of electrons are not represented as separate entities, but as interactions sites representing either the potential of all electrons in a bond or both electrons in a lone pair.\\

\begin{figure}[ht!]
 \begin{center}
 \resizebox{1.0\columnwidth}{!}{\includegraphics[scale=0.5,clip,angle=0]{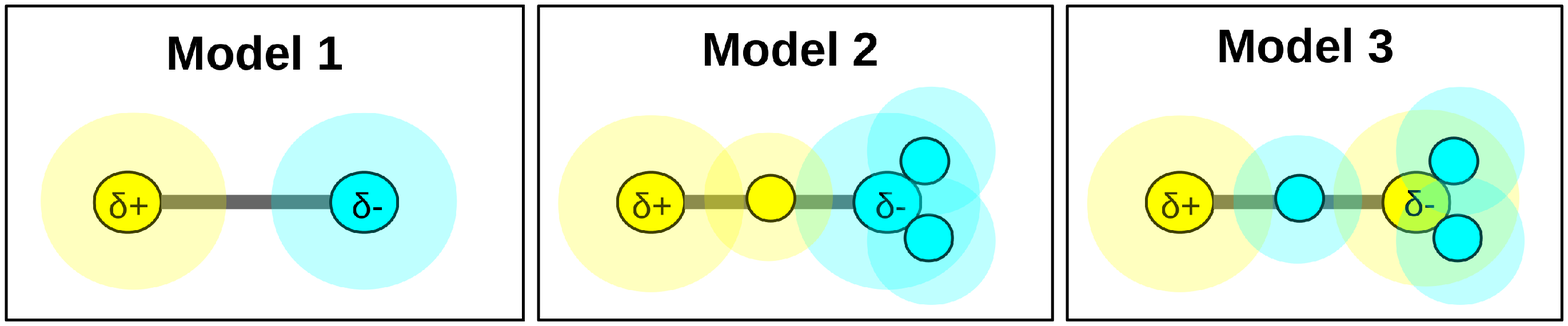}}
 \caption{\small Schematic representation of the three interaction models for a hypothetical molecule with an atom of lower electronegativity on the left ($\delta+$) and an atom of higher electronegativity ($\delta-$) and two lone pairs on the right. Yellow color indicates positive charge, blue indicates negative charge.}
\label{fig:models-scheme}
\end{center}
\end{figure}

\noindent Each of the three interaction models is defined by specific potential functions and by a scheme for representing 
the charge distribution in a molecule used for parametrization. One of the problems with the introduction of 
additional interaction sites is the fact that the parameter space is increased as more sites are added and  therefore 
requires a more expensive parametrization procedure. In order to avoid this and keep the parametrization effort for all 
models equal, the three models are constructed such that all models require the same number of adjustable parameters and each 
model only requires one adjustable parameter for the electrostatic interactions per molecule. A schematic representation of the charge distribution
scheme in the three models is shown in Figure \ref{fig:models-scheme}. The equal number of parameters is achieved by 
assigning generic values to a number of model parameters which are treated as constant for all molecules. As a 
consequence all models are equally complex as far as the parametrization is concerned and an increase in accuracy does 
not translate into a higher parametrization effort. The parametrization of each model is described in 
detail in Section 'Methods'. The three models can be summarized by the following properties:
\begin{itemize}
\item[Model 1] is the model closest to conventional force fields as it just contains atom-centered potential functions. 
The charge distribution is represented by atom-centered potentials parametrized based on the electronegativity difference between each pair of atoms in a bond. 
\item[Model 2] contains additional interaction sites which are either defined as bond electrons
potentials placed between two atoms or as lone pair electron potentials at a given distance of an atom. The interaction 
potential consists of spherical electrostatic potential functions on the atoms and the
additional interaction sites. In addition to the electrostatic potential, the different sites interact via Pauli repulsion potentials. The electron potential parameters are determined based on the electronegativity differences 
as in Model 1, with bond potential sites sharing the charge of the less electronegative atom in the bond and 
lone pair potentials sharing the charge of their corresponding atom.
\item[Model 3] contains additional charge sites with positions defined as for Model 2. In contrast to Model 2 the concept 
of Model 3 is not just to have additional interaction sites sharing the charge of the atoms, but to treat the atoms centers as positively charged atom cores and the additional sites as valence electron potentials. The atom cores are represented by point charges whereas Gaussian distributions are used for the electron potentials. The atom and electron potential parameters are assigned based on the valence electron structure of 
the molecules with positive charge centered at the atom cores and negative charge centered on the electron potentials. 
\end{itemize}

\noindent The choice of potential functions to describe electrostatic interactions is an important point to evaluate. 
In conventional atomistic force fields electrostatics are represented by point charges, whereas in eFF they are described 
by spherical Gaussian distributions. In order to compare the electrostatics based on spherical Gaussian distributions to 
conventional point charges, each of the three models is evaluated in two versions, once with point charges and once with 
spherical Gaussian functions. In the point charge version Model 1 just contains conventional force field functions and 
can therefore be used for comparison. All interaction potential functions are described in detail in Section 'Methods'.\\

\noindent The properties of the three models will be explored in the following using a test 
set of 12 small molecules. The 12 molecules are chosen such that the set contains different types of valence electron
structures, different chemical elements as well as aromatic molecules with highly delocalized electrons. 
First, the accuracy of the electrostatic potential will be evaluated for the 
different models. Each model will be evaluated as a charge distribution model and as a point charge model. Based on the results, the accuracy of intermolecular interaction energies will be compared for the different models using 
a subset of 8 molecules.\\

\section{Methods}
\label{sec:methods}

\subsection{Interaction potential functions}

\subsubsection{Gaussian charge distribution models}

\noindent For {\bf Model 1}, the nonbonded interaction $V_{M1}$ consists of a spherical Gaussian
function for the electrostatic interaction and an LJ term representing dispersion and Pauli
repulsion.

\begin{align}
V_{M1}(\vec{r}) = \sum_{nonbonded}\Big[q_i q_j\int\frac{|\psi_i|^2|\psi_j|^2}{R_{ij}}dr + \epsilon(\frac{\sigma_j^{12}}{R_{ij}^{12}}-2\frac{\sigma_{ij}^6}{R_{ij}^6})\Big],
\label{model1_eq}
\end{align}

\noindent with $R_{ij}$ being the interatomic distance, $q_i$ being the charge centered on atom 
$i$,  $\psi_i$ being the Gaussian distribution on atom $i$, $\sigma$ and $\epsilon$ being the LJ radius and well depth.
The integrals over the Gaussian distributions required for the electrostatic 
interactions between two atoms $i$ and $j$ can be evaluated as a function of the distance $R_{ij}$ using the error functions since

\begin{equation}
\int\frac{|\psi_i|^2|\psi_j|^2}{R_{ij}}dr=\frac{1}{R_{ij}}Erf\frac{\sqrt{2}R_{ij}}{\sqrt{\omega_i^2+\omega_j^2}}. 
\end{equation}

\noindent with $\omega$ being the width of the Gaussian distribution. In analogy, the interaction 
between a point charge and a Gaussian distribution can be expressed as 

\begin{equation} 
\int\frac{|\psi_j|^2}{R_{ij}}dr=\frac{1}{R_{ij}}Erf\frac{\sqrt{2}R_{ij}}{\omega_j},
\end{equation}

\noindent Based on these expressions, 
the integrals required for the calculation of the interaction potential can be implemented in a 
computationally inexpensive way using numerically efficient methods to evaluate the error function,
 as used in the eFF implementation.\cite{Jaramillo-BoteroGoddard2011}\\

\noindent For {\bf Model 2}, the nonbonded interaction of the atoms $V_{M2\_at}$ has the 
same terms as $V_{M1}$, whereas for the electron potential sites $V_{M2\_elp}$, the LJ potential is 
replaced by a repulsive potential for simplicity in order to reduce the interaction to the minimal requirements 
for an electron potential.

\begin{align}
V_{M2\_at}(\vec{r}) = \sum_{nonbonded}\Big[q_i q_j\int\frac{|\psi_i|^2|\psi_j|^2}{R_{ij}}dr + \epsilon(\frac{\sigma_j^{12}}{R_{ij}^{12}}-2\frac{\sigma_{ij}^6}{R_{ij}^6})\Big],
\end{align}

\begin{align}
V_{M2\_elp}(\vec{r}) = \sum_{nonbonded}\Big[q_i q_j\int\frac{|\psi_i|^2|\psi_j|^2}{R_{ij}}dr + \epsilon(\frac{\sigma_{ij}}{R_{ij}})^{6}\Big],
\label{model2_elp_eq}
\end{align}

\noindent with the $\epsilon$ parameter being identical for both potential parts.\\

\noindent For {\bf Model 3}, the nonbonded interaction of the atoms $V_{M3\_at}$ is described by 
point charges instead of Gaussian distributions. 

\begin{align}
V_{M3\_at}(\vec{r}) = \sum_{at\_at}\Big[\frac{q_iq_j}{R_{ij}} + \epsilon(\frac{\sigma_j^{12}}{R_{ij}^{12}}-2\frac{\sigma_{ij}^6}{R_{ij}^6})\Big],
\end{align}

\noindent For the interaction between electron potential sites $V_{M3\_elp\_elp}$, the potential 
function is identical to $V_{M2\_elp}$ (Equation \ref{model2_elp_eq}). For the interaction between
atom cores and electron potentials $V_{M3\_at\_elp}$, interactions between Gaussian distributions 
and point charges have to be evaluated.

\begin{align}
V_{M3\_at\_elp}(\vec{r}) = \sum_{at\_elec}\Big[q_i q_j\int\frac{|\psi_j|^2}{R_{ij}}dr + \epsilon(\frac{\sigma_{ij}}{R_{ij}})^{6}\Big],
\end{align}

\noindent with the $\epsilon$ parameter again being identical for both potential parts.\\

\subsubsection{Point charge models}
\noindent For each of the three interaction models, a point charge model is evaluated in addition. In these models the electrostatic
interactions $V_{elec}$ based on Gaussian functions

\begin{align}
V_{elec}=q_i q_j\int\frac{|\psi_i|^2|\psi_j|^2}{R_{ij}}dr
\end{align}

\noindent are replaced on all interactions sites by 

\begin{align}
V_{elec}=\frac{q_iq_j}{R_{ij}}
\end{align}

\noindent In the case of Model 1, the interaction potential is then identical to conventional force field functions.\\

\subsection{Parametrization of interaction models}

\noindent As explained above, the number of adjustable parameters is identical for all three models and there is only one adjustable
parameters for the electrostatic potential per molecule. This is achieved as follows: The positions
of the additional interaction sites are determined based on the molecular structure 
and the Slater radii $S$ \cite{Slater64} of the atoms. Bond electron potentials are simply 
placed at the geometric center of each bond. The position of the lone pair potential of atom $i$ 
is determined by constraining it to a generic distance of $\frac{3}{4}S_i$ from the atom and 
minimizing the interaction energy between all electron potentials in the molecule given in 
Equation \ref{model2_elp_eq}. For the electrostatics the parameter space is reduced by either 
considering charge transfer between atoms due to electronegativity differences or by using the 
valence electron structure of the molecule. For the $\sigma$ and $\omega$ parameters, additional interaction sites in
principle require additional parameters. For $\omega$ this is avoided here by using only one adjustable parameters for
all atoms and electron potential sites based on initial parameter values for atom radii and generic values for
electron potentials. For $\sigma$ additional parametrization of the electron potential sites is avoided by using a
generic value of $\sigma_{elp}=1.0$ \AA\ on all electron potential sites.\\

\noindent The parametrization is carried out separately for the electrostatic potential surface 
(EPS) and intermolecular interactions since the optimal parameters are not identical in both
cases.\cite{KramerMeuwlyJCTC2013} This is on one hand due to polarization effects, on the other
hand due to error compensation. For the EPS partial charges need to be determined for all models and the width of
the Gaussian distribution $\omega$ needs to be determined for the Gaussian charge distribution
models. For the intermolecular interactions the width of the $\sigma$ and $\epsilon$ parameters of the LJ potential
and the repulsive potential for the electron potential sites needs to be determined in addition. The $\omega$ and
$\sigma$ parameters are in principle related as both of them describe the width of the electron density distribution. However, since they are parametrized differently and used in different evaluations two different symbols are used for clarity.\\

\subsubsection{Charges on atoms and electron potentials}

\noindent For {\bf Models 1 and 2} partial charges are determined
based on Pauling electronegativities\cite{Pauling32} $P$. For each pair 
of atoms $i$ and $j$ in a bond, an electronegativity difference coefficient 
$\delta_{bond}=1.0-\frac{P_i}{P_i+P_j}$ is determined. For atom $i$ forming $Nb$ single-, 
double-, or triple-bonds with $Zb_{n}$ binding electrons, electronegativity based charges
$q_i$ are then determined as 

\begin{equation}
q_i=\zeta \sum_{n=1}^{Nb}(0.5-\delta_n)Zb_n,
\end{equation}

\noindent with $\zeta$ being an adjustable molecular parameter which is determined either by 
fitting to the {\it ab initio} EPS or intermolecular interaction energies. For Model 1 these 
charges are used directly, whereas for Model 2 the atom charge population is distributed equally 
among all sites assigned to the atom. Lone pair electron potential sites are assigned to 
their corresponding atom, whereas bond electron potential sites are assigned to the atom in the 
bond with the smaller electronegativity. In the case of 
equal electronegativity the sites are assigned to both atoms equally.\\

\noindent {\bf Models 3}: The initial bond charges $z_i$ for this model are given by the bond order, i.e. $z_i=-2.0$ for single bonds, $z_i=-4.0$ for double bonds
and $z_i=-6.0$ for triple bonds. In the case of delocalized bonds the charge is distributed 
equally amongst all sites sharing it, e.g. for all benzene C-C bonds $z_i=-3.0$. For lone pair 
potentials $z_i=-2.0$. For each atom core the basic charge $z_i$ is determined as the charge of the nucleus minus the number of valence electrons. The initial charge $z_i$ is however only the correct atom core charge in cases where no charge transfer takes place between 
atoms and electrons are localized entirely on the on the electron potential sites. In order to 
correct for these shortcomings the final charges are determined as follows: for atom $i$ forming 
$Nb$ bonds with $Zb_{n}$ binding electrons, the atom charge $q_i$ is determined as

\begin{equation}
q_i=\zeta z_i-\sum_{n=1}^{Nb}(0.5-\delta_n)Zb_n, 
\end{equation}

\noindent with $\zeta$ being again an adjustable molecular parameter which is applied to the 
charges on all sites in a molecule and determined either by fitting to the {\it ab initio} EPS or 
to intermolecular interaction energies. In contrast to Models 1 and 2, the charge transfer 
between atoms is not affected by the $\zeta$-parameter in Model 3. The effect of a small
$\zeta$-value is equivalent to having parts of the valence electron density localized on the atom 
core, i.e. the charge separation is smaller.\\

\subsubsection{Width of Gaussian distributions and LJ potential parameters}

\noindent {\bf Width $\omega$ of Gaussian distributions:} 
As initial value for $\omega$ the Slater radius $S$ \cite{Slater64} is used for the atoms and 
a generic initial value of $\omega$=1.0 \AA\ for all electron potential sites. 
Based on these initial values, the final width $\omega_i$ is determined using again a single 
scaling parameter $\upsilon$ per molecule.\\
 
\noindent {\bf Radius $\sigma$ and $\epsilon$ for LJ and repulsive potentials:} On the atoms, 
empirical values derived from X-ray diffraction data \cite{Bondi64} 
are used for $\sigma$ except in the case of hydrogen where this value is generally too large and 
therefore replaced by the Slater radius\cite{Slater64} of $\sigma$=0.25 \AA\/. For  
$\epsilon$ a single adjustable parameter is again used which is determined by fitting to the 
intermolecular interaction energies.\\

\subsubsection{Adjustable parameter fitting}

\noindent As the number of adjustable parameters is kept small, fitting parameters to either the EPS or 
intermolecular interaction energies is simple. For fitting the point charge models to the 
EPS there is only one parameter $\zeta_m$ to be determined for each molecule $m$. This is done by 
calculating the average potential energy difference $\bar{\Delta}_{Epot}$ between the {\it ab initio} 
EPS outside the Slater radii of each atom on a 3-dimensional grid of 20 \AA\ side length centered about the geometric center of the molecule, 
with a distance of 1.0 \AA\ between grid points. The molecular parameter $\zeta_m$ is determined as 
$\zeta_m=ArgMin(\bar{\Delta}_{Epot}(\zeta))$ in the range $\zeta=[0.0,4.0]$.\\

\noindent For fitting the Gaussian charge distribution models to the EPS two parameters, $\zeta_m$ and 
$\upsilon_m$ need to be determined. This is done in two steps: first $\upsilon_m=ArgMin(\bar{\Delta}_{Epot}(\upsilon)| \zeta_m)$  is determined in the 
range $\upsilon=[0.0,8.0]$ for $\zeta_m=\{\zeta_{mPC},\frac{5}{4}\zeta_{mPC}\}$, with $\zeta_{mPC}$ being the $\zeta$-parameter determined 
for the point charge model. In the second step $\zeta_m$ is determined as $\zeta_m=ArgMin(\bar{\Delta}_{Epot}(\zeta)| \upsilon_m)$ in the
range $\zeta=[0.0,4.0]$. The parameters for fitting all models to the EPS are given in in Table \ref{tb:params1}.\\

\begin{table}[h] 
\caption{Molecular parameters fitted to the EPS for all three interaction models in their point 
charge and Gaussian charge distribution version. Parameter values of 0.0 indicate that 
$\bar{\Delta}_{Epot}$ is smallest if the corresponding energy term is omitted. 
{\footnotesize $^a$ For PH$_3$ Mulliken electronegativities\cite{Mulliken34} are used instead of 
Pauling electronegativities due to the small Pauling electronegativity difference between the  
atoms.}}
\begin{center}
\begin{tabular}{l||c|c|c||c|c|c|c|c|c}   
   & \multicolumn{3}{c||}{point charge models} & \multicolumn{6}{c}{Gaussian distribution models} \\
\hline 
   & Model 1 & Model 2 & Model 3 & \multicolumn{2}{c|}{Model 1} & \multicolumn{2}{c|}{Model 2} & \multicolumn{2}{c}{Model 3} \\
\hline   
       & $\zeta$ & $\zeta$ & $\zeta$ & $\zeta$ & $\upsilon$ & $\zeta$ & $\upsilon$ & $\zeta$ & $\upsilon$  \\
\hline 
\hline
H$_2$O & 1.7 & 1.6 & 0.15 & 1.7 & 0.9 & 1.6 & 1.0 & 0.15 & 1.2 \\
\hline
NH$_3$ & 1.9 & 1.3 & 0.15 & 1.9 & 1.0 & 1.3 & 1.1 & 0.15 & 1.4 \\
\hline
CH$_4$ & 1.9 & 3.4 & 0.1 & 2.1 & 1.0 & 3.7 & 0.9 & 0.15 & 1.6 \\
\hline
H$_2$S & 1.1 & 1.0 & 0.01 & 1.1 & 1.6 & 1.0 & 2.1 & 0.015 & 3.5 \\
\hline
PH$_3$$^a$ & 0.9 & 0.8 & 0.0 & 1.0 & 0.9 & 0.8 & 2.2 & 0.01 & 4.4 \\
\hline
CO$_2$ & 2.2 & 2.1 & 0.1 & 2.2 & 0.7 & 2.1 & 0.9 & 0.15 & 0.6 \\
\hline
ethanol & 0.9 & 0.9 & 0.05 & 0.9 & 1.1 & 0.9 & 1.1 & 0.05 & 1.8 \\
\hline
CH$_5$N & 1.2 & 0.9 & 0.15 & 1.2 & 1.4 & 0.9 & 1.6 & 0.15 & 1.4 \\
\hline
OCH$_2$ & 1.7 & 1.7 & 0.25 & 1.7 & 0.7 & 1.7 & 0.9 & 0.25 & 1.2 \\
\hline
benzene & 1.7 & 2.1 & 0.1 & 1.7 & 1.3 & 2.2 & 1.5 & 0.1 & 1.7 \\
\hline
pyrrole & 0.0 & 0.0 & 0.3 & 0.0 & 0.0 & 0.0 & 0.0 & 0.3 & 1.3 \\
\hline
thiophene & 1.4 & 0.9 & 0.15 & 1.4 & 1.2 & 0.9 & 0.9 & 0.15 & 1.7 \\
\hline
glycine & 1.4 & 0.5 & 0.0 & 1.5 & 1.4 & 0.5 & 0.9 & 0.01 & 3.6 \\
\hline
\end{tabular}	            
\end{center}
\label{tb:params1}
\end{table}

\noindent For fitting the point charge models to intermolecular interaction energies there are also two parameters to 
determine, $\zeta_m$ and $\epsilon_m$. This is done by calculating the average potential energy difference 
$\bar{\Delta}_{E\_inter}$ between a set of 26 {\it ab initio} intermolecular interaction energies and the interaction 
energy calculated from each model. (For details to the {\it ab
initio} calculations see next section). In the first step $\epsilon_m$ is determined as 
$\epsilon_m=ArgMin(\bar{\Delta}_{E\_inter}(\epsilon)| \zeta_m)$ in the range $\epsilon=[0.0,2.0]$ for 
$\zeta_m=\{0.1,1.0\}$. In the second step $\zeta_m=ArgMin(\bar{\Delta}_{E\_inter}(\zeta)|
\epsilon_m)$ is determined in the range 
$\zeta=[0.0,4.0]$. The parameters for fitting all models to interaction energies are given in in Table \ref{tb:params2}.\\

\begin{table}[h] 
\caption{Molecular parameters fitted to the intermolecular interaction energies for all three interaction models. Parameter values of 
0.0 indicate that $\bar{\Delta}_{E\_inter}$ is smallest if the corresponding energy term is omitted.}
\begin{center}
\begin{tabular}{l||c|c|c|c|c|c}   
    & \multicolumn{2}{c|}{Model 1} & \multicolumn{2}{c|}{Model 2} & \multicolumn{2}{c}{Model 3} \\
\hline   
     & $\zeta$ & $\epsilon$ & $\zeta$ & $\epsilon$ & $\zeta$ & $\epsilon$  \\
\hline 
\hline
H$_2$O & 1.75 & 0.51 & 1.0 & 0.13 & 0.0085  & 0.09 \\
\hline
NH$_3$ & 1.0 & 0.13 & 1.1 & 0.31 & 0.009 & 0.29 \\
\hline
CH$_4$ & 0.0 & 0.57 & 0.1 & 0.3 & 0.0 & 0.66 \\
\hline
H$_2$S & 0.0 & 0.71 & 0.7 & 0.16 & 0.0045 & 0.16 \\
\hline
PH$_3$$^a$ & 1.2 & 1.23 & 0.45 & 1.5 & 0.0 & 1.38 \\
\hline
CO$_2$ & 1.55 & 0.16 & 1.55 & 0.03 &  0.0045 & 0.03 \\
\hline
CH$_5$N & 0.1 & 0.83 & 0.1 & 0.85 & 0.0005 & 0.55 \\
\hline
OCH$_2$ & 1.2 & 0.71 & 1.0 & 0.12 & 0.0006 & 0.0 \\
\hline
\end{tabular}	            
\end{center}
\label{tb:params2}
\end{table}

\subsection{Electronic structure calculations and distributed multipole moments}

\noindent {\bf EPS calculations:} Electrostatic potential surfaces have been calculated for all test molecule 
using ORCA\cite{ORCA}. As a reference for parameter fitting, density functional theory was used with the B3LYP 
functional\cite{Becke93} and an aug-cc-pVTZ basis set.\cite{Dun89}. The EPS was evaluated outside the Slater 
radii\cite{Slater64} of the molecule on a 3-dimensional grid of 20 \AA\ side length centered about the geometric 
center of the molecule, with a distance of 1.0 \AA\ between grid points.
For the evaluations of the accuracy, a reference EPS was calculated at the MP2/aug-cc-pVTZ level of theory.
The reference EPS was evaluated on a 3-dimensional grid with gridpoints placed at radial distances of 
up to 10 \AA\ about any atom in the molecule, with a distance of 0.25 \AA\ between gridpoints. This 
choice of gridpoints is due to the fact that the evaluation is carried out for point segments within 
selected distance ranges of any atom as explained in the Section 'Results'.
For comparison Mulliken charges\cite{Mulliken55} and distributed multipole moments\cite{Stone96} were 
calculated based on B3LYP/aug-cc-pVTZ calculations carried out with GAUSSIAN.\cite{Gaussian03}
Distributed multipole moments were obtained based on the electron density 
distributions using GDMA.\cite{GDMA05}\\

\noindent {\bf Intermolecular interaction energies:} Intermolecular interaction energies were calculated at the
MP2/aug-cc-pVTZ level of theory for 26 dimer geometries of the eight molecules selected for the
evalulation using ORCA.\cite{ORCA} Counterpoise corrections were used to account for the basis
set superposition error.\cite{SimonDannenberg96} For each of the selected molecules, the dimer geometry 
was first optimized. Starting from the optimized geometry, 25 new dimer geometries were generated 
by random translation and rotation of the two monomers. In order to avoid very unfavorable geometries which 
are not of interest for the evaluation as they are unlikely to be observed, geometries with atom 
distances $<1.0$ \AA\ were rejected.\\

\vspace{3.5cm}

\section{Results}

\subsection{Comparison of electrostatic potential surfaces}

\begin{figure}[ht!]
 \begin{center}
 \resizebox{0.7\columnwidth}{!}{\includegraphics[scale=0.5,clip,angle=0]{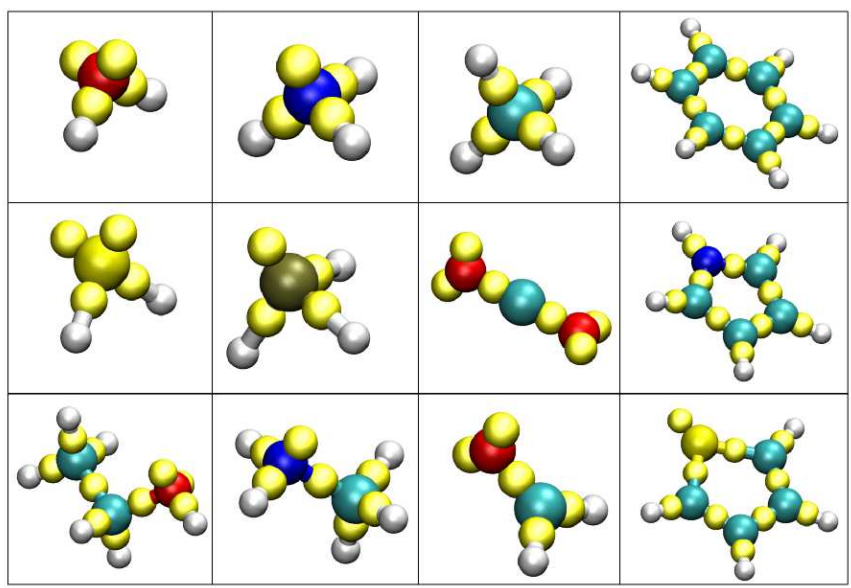}}
 \caption{\small Molecules with optimized electron potential coordinates (yellow spheres). First row: H$_2$O, NH$_3$, methane 
 benzene. Second row: H$_2$S, PH$_3$, CO$_2$, pyrrole. Third row: ethanol, methylamine, formaldehyde, thiophene.}
\label{fig:molecules}
\end{center}
\end{figure}

\noindent As a first step for all evaluations, the coordinates of the electron potential sites need to be determined. The coordinates of the bond electron potentials are calculated based on the atom coordinates, while lone pair electron potential coordinates are obtained by minimizing the interaction energy of all electron potentials in the molecule using the nonbonded 
interaction potentials given in Section 2.1. The resulting coordinates for the 12 molecules 
used in the following are shown in Figure \ref{fig:molecules}.\\

\begin{figure}[ht!]
 \begin{center}
 \resizebox{0.85\columnwidth}{!}{\includegraphics[scale=0.5,clip,angle=0]{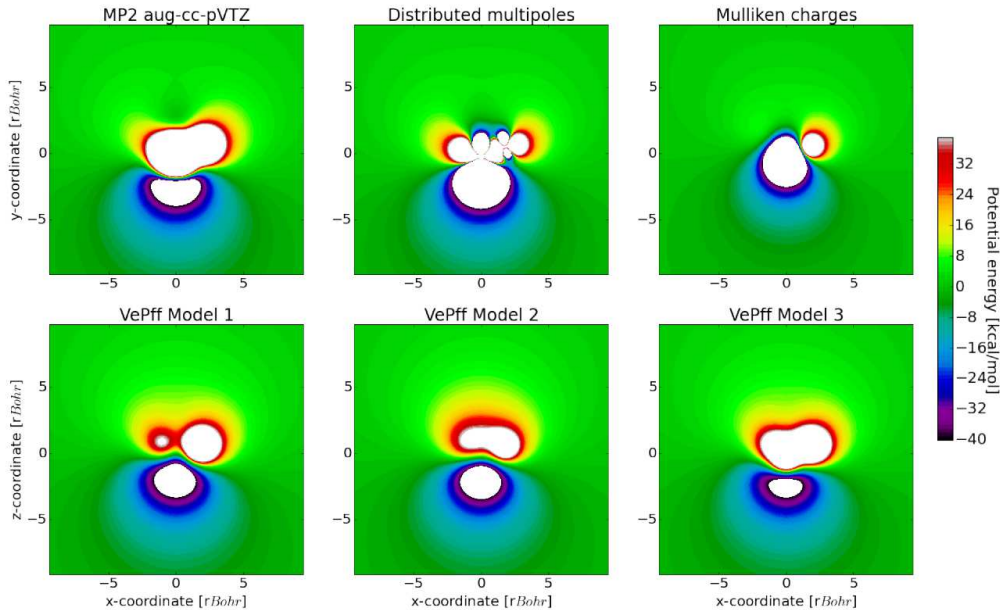}}
 \caption{\small Electrostatic potential surfaces for NH$_3$. Upper row: comparison to   MP2/aug-cc-pVTZ, DMA and Mulliken charges. Lower row: EPS of the three valence electron potential force field (VePff) models. White areas: energies outside the range of $\pm 40$ kcal/mol.}
\label{fig:NH3}
\end{center}
\end{figure}

\begin{figure}[ht!]
 \begin{center}
 \resizebox{0.85\columnwidth}{!}{\includegraphics[scale=0.5,clip,angle=0]{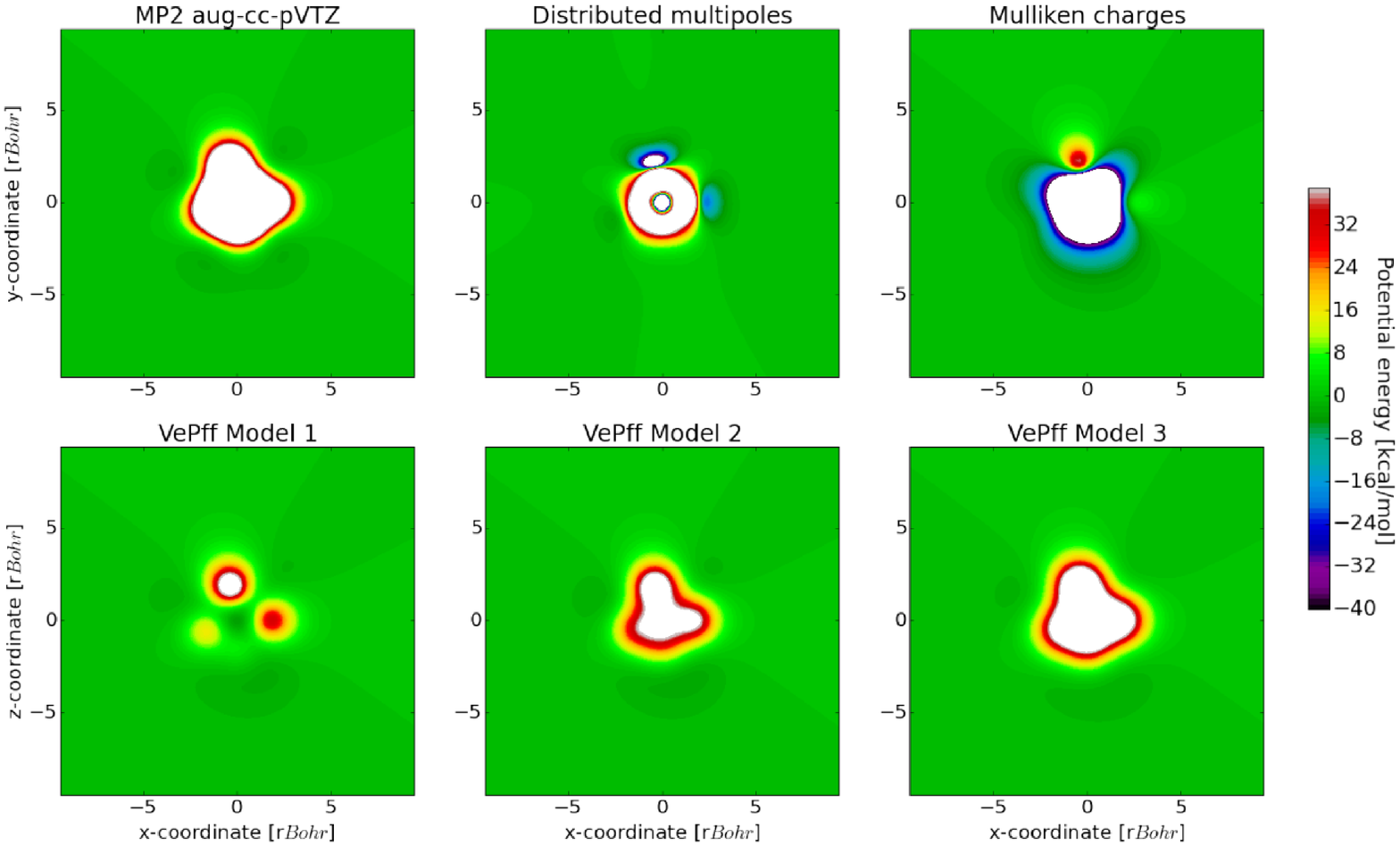}}
 \caption{\small Electrostatic potential surfaces for CH$_4$. Upper row: comparison to   MP2/aug-cc-pVTZ, DMA and Mulliken charges. Lower row: EPS of the three valence electron potential force field (VePff) models. White areas: energies outside the range of $\pm 40$ kcal/mol.}
\label{fig:CH4}
\end{center}
\end{figure}

\begin{figure}[ht!]
 \begin{center}
 \resizebox{0.85\columnwidth}{!}{\includegraphics[scale=0.5,clip,angle=0]{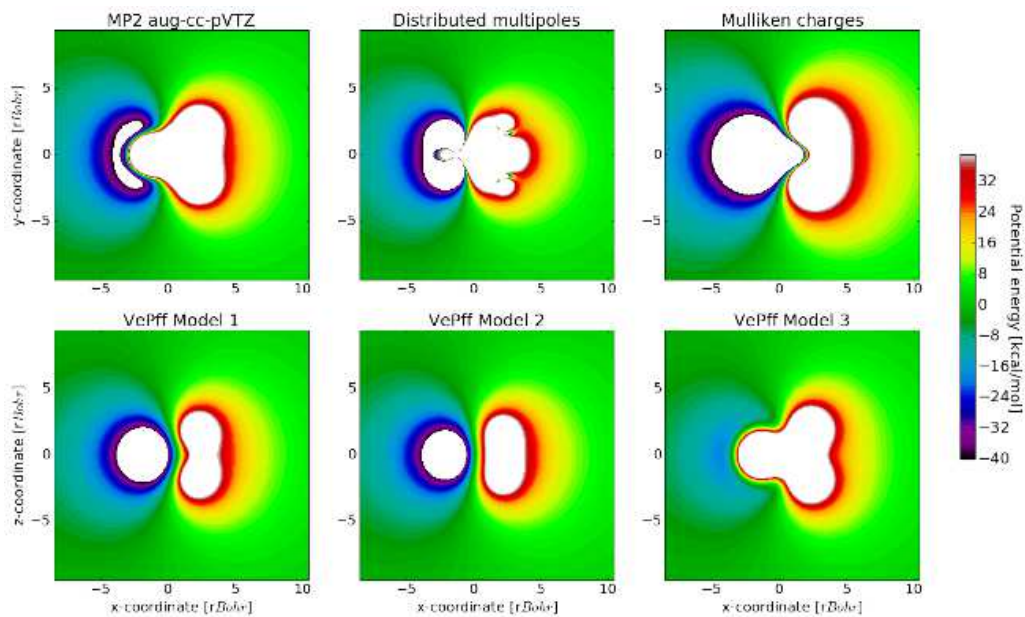}}
 \caption{\small Electrostatic potential surfaces for OCH$_2$. Upper row: comparison to   MP2/aug-cc-pVTZ, DMA and Mulliken charges. Lower row: EPS of the three valence electron potential force field (VePff) models. White areas: energies outside the range of $\pm 40$ kcal/mol.}
\label{fig:OCH2}
\end{center}
\end{figure}

\noindent The differences between the EPS of the three models, {\it ab initio} calculations, distributed multipoles and Mulliken charges are first illustrated for three molecules, NH$_3$, CH$_4$ and formaldehyde (OCH$_2$). The EPS of all models is calculated on a two-dimensional grid and is shown in Figures \ref{fig:NH3}, \ref{fig:CH4} and
\ref{fig:OCH2}. Comparison of the potential energy surfaces shows that that the EPS of Model 3 is qualitatively the most similar to the {\it ab initio} EPS calculated at the MP2/aug-cc-pVTZ level of theory. The largest differences to {\it ab intio} are observed for Mulliken charges.
Models 1 and 2 have similar potential surfaces, with Model 2 appearing as slightly 
better than Model 1 overall. For distributed multipoles it can clearly be seen that the EPS is becoming more accurate as the distance to the atoms becomes larger. This is due to the fact that the distributed multipole expansion converges at a given radius from the 
multipole sites which varies for different molecules. In addition to the convergences of the multipole expansion with distance there is a convergence of the expansion as a function of 
the highest multipole rank to consider. For this and the following evaluations, multipole 
expansions are truncated at rank 2 (quadrupole). This rank has been found to provide a good trade-off between convergence and the computational effort to calculate multipolar energies 
and interactions.\cite{JMolM09}\\

\noindent The comparison of the EPS above is obviously just qualitative, but it illustrates the differences between the {\it ab inito} EPS and the different 
models. Most importantly it shows that there are systematic differences between the three models that are observed for different molecules.  Furthermore the accuracy of the different models varies as a function of the distance to the atoms. Models which are more similar to 
{\it ab initio} in the short range may perform badly at longer distances and vice versa, therefore for the quantitative evaluation of the EPS accuracy in the next section, the distance dependence will be considered.\\

\clearpage

\subsection{Evaluation of electrostatic potential accuracy}

\begin{figure}[ht!]
 \begin{center}
 \resizebox{0.75\columnwidth}{!}{\includegraphics[scale=0.5,clip,angle=0]{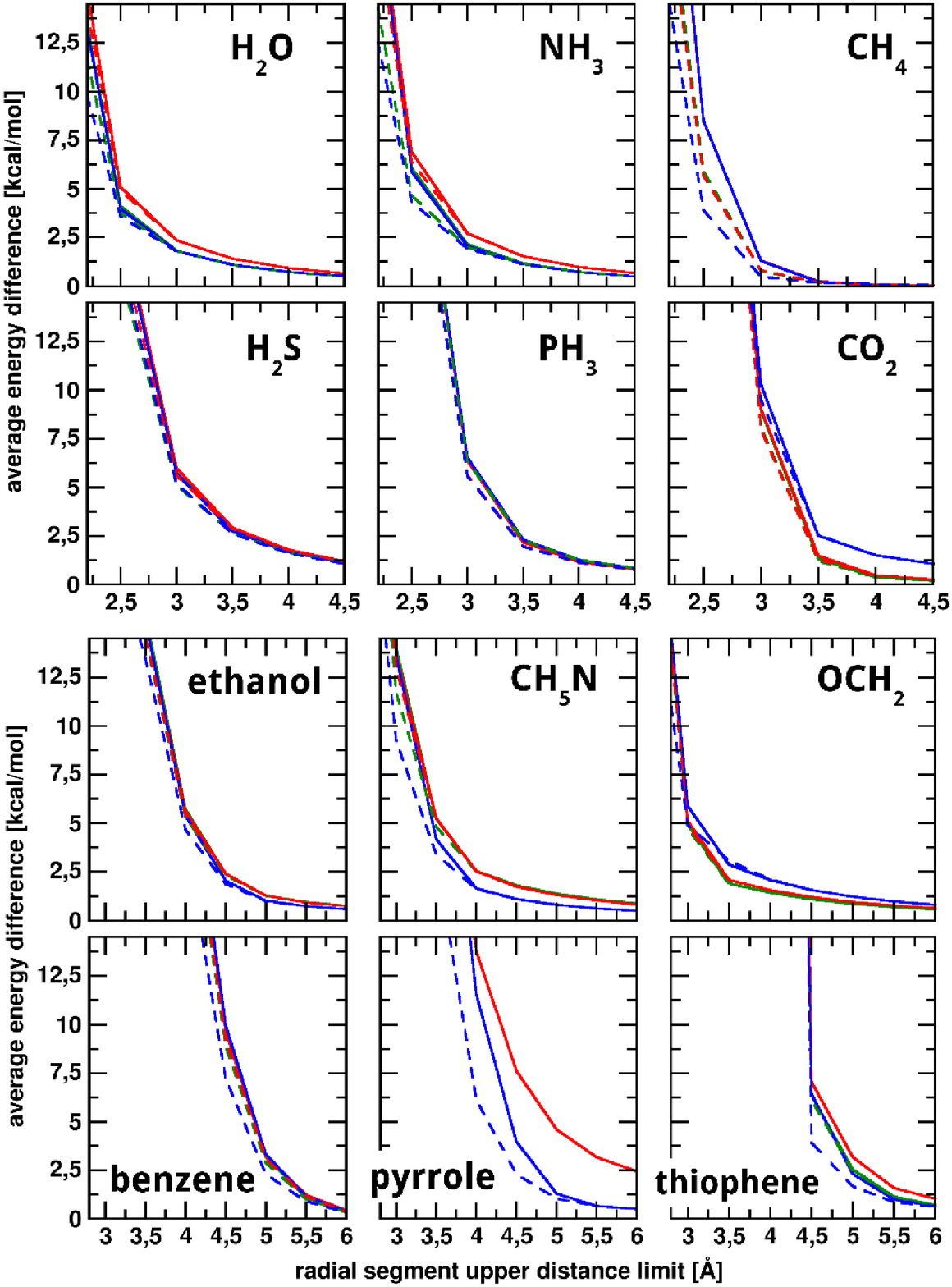}}
 \caption{\small Accuracy evaluation for Model 1 (green), Model 2 (red) and 
 Model 3 (blue) in their point charge (solid lines) and Gaussian distribution version (dashed lines).  The differences between each model and the MP2/aug-cc-pVTZ EPS are evaluated for segments of increasing distance to the atoms. The upper six molecules contain only 
 two atom types, the lower six molecules more than two atom types.}
\label{fig:eval-models}
\end{center}
\end{figure}

\begin{figure}[ht!]
 \begin{center}
 \resizebox{0.75\columnwidth}{!}{\includegraphics[scale=0.5,clip,angle=0]{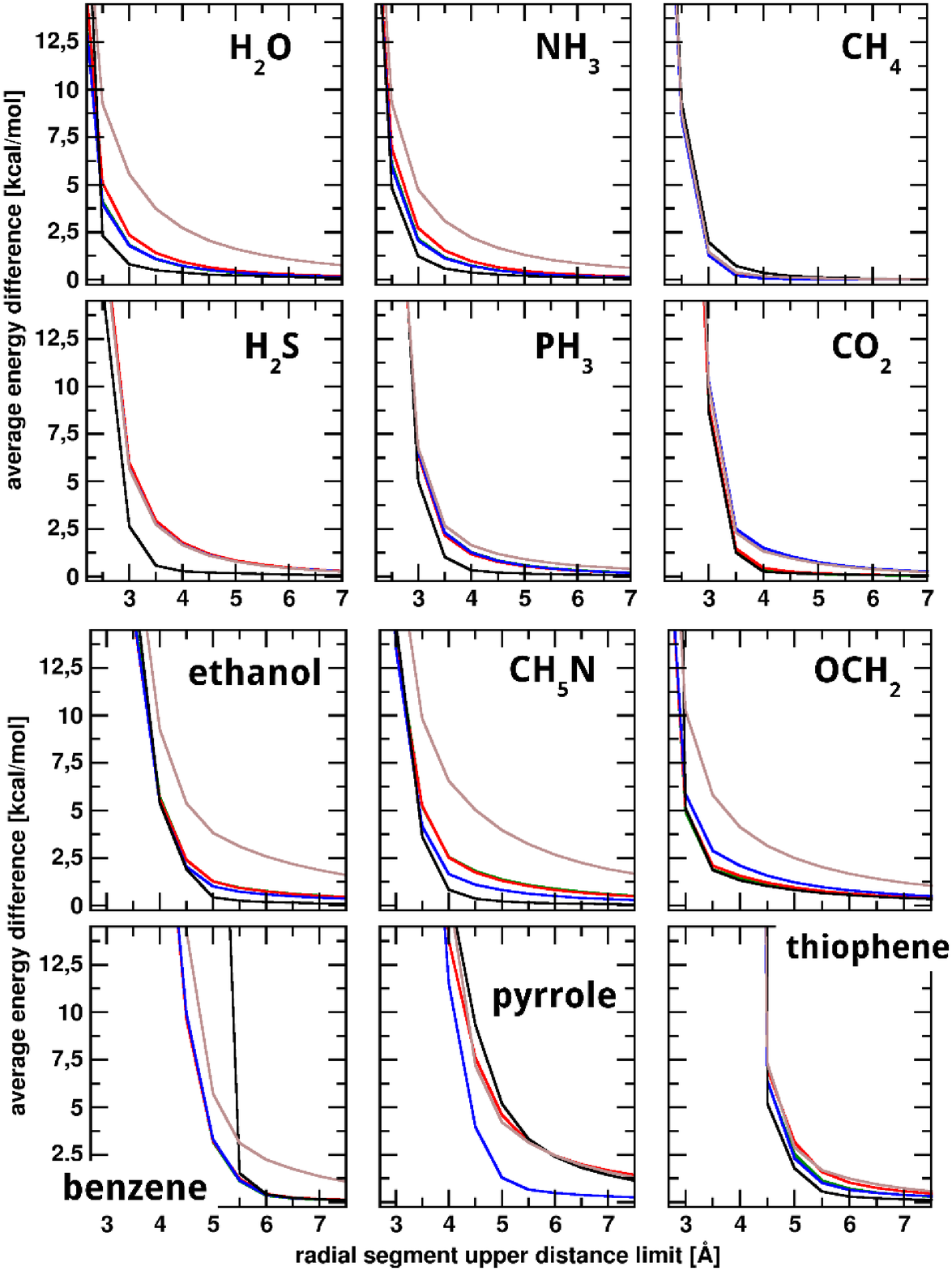}}
 \caption{\small Comparison of the accuracy of point charge Model 1 (green), Model 2 (red) and Model 3 (blue) to DMA (black) and Mulliken charges (brown). The differences between each model and the MP2/aug-cc-pVTZ EPS are evaluated for segments of increasing distance to the atoms. The upper six molecules contain only two atom types, the lower six molecules more than two atom types.}
\label{fig:compare-models}
\end{center}
\end{figure}

\noindent In order to quantitatively compare the EPS accuracy, the differences between the electrostatic potential of different models and the {\it ab initio} EPS calculated at the MP2/aug-cc-pVTZ level of theory are evaluated on a 3-dimensional grid centered at the geometric center of the molecules. The evaluation is carried out for segments of points within
increasing distance ranges of any atom in the molecule. This distance dependent evaluation allows to compare not only the overall accuracy of each model, but also to assess the 
distance range in which the errors occur. As the electrostatic potential is higher at short distances, the errors are in general also higher in the near range.
In Figure \ref{fig:eval-models} the results are compared between the three models in their point charge version and in their Gaussian distribution version. In 
Figure \ref{fig:compare-models} the point charge version of the three models is compared to Mulliken charges and distributed multipoles. For clarity additional representation of this comparison for each model separately are shown in the Supplementary material (SI Figures 1-3).\\

\noindent All interaction models including DMA and Mulliken charges have been parametrized at the B3LYP/aug-cc-pVTZ level of theory, therefore the evaluations here also assess the transferability between a computationally less expensive method used for parametrization and a computationally more expensive and more accurate reference. The molecules are divided into two groups of six molecules where the upper group of six molecules in both Figures shows molecules with only two atom types, whereas the lower group contains more than two atom types. This distinction is important due to the fact that for molecules with only two atom types, the parameter space of the electrostatic parameter $\zeta$ contains a value 
corresponding to Mulliken charges. Therefore if these molecules are more accurate than Mulliken charges this is mainly due probing a larger parameter space, as well as due to the transferability between the DFT EPS used for fitting and the MP2 reference calculation. For 
the molecules with more than two atom types in contrast, the parameter space does not 
necessarily contain Mulliken charges, therefore higher accuracy than Mulliken charges implies a real advantage of this model and parametrization.\\

\noindent Comparison of the evaluations on all 12 molecules shows that in general Mulliken charges are the least accurate method, DMA is most accurate in the longer range, whereas in the short range its equally or less accurate than the three VePff models. Models 1 and 2 perform very similar, therefore the green lines representing Model 1 are mostly covered by the the red lines representing Model 2. Model 3 performs better than Models 1 and 2 except for CO$_2$ and and formaldehyde. The Gaussian distribution models are systematically more accurate than the point charge models, however the gain in accuracy is mostly small and mainly relevant in the short range. Due to this result, the Gaussian charge distribution models will not be evaluated for intermolecular interaction energies in the next section as the differences between 
different models are found here to be substantially more important than differences between point charges and Gaussian distributions. Furthermore intermolecular interaction energies are dominated by repulsive interaction potentials in the short range, so most likely the accuracy of the repulsive energy terms is more important than the additional accuracy due to the use of Gaussian functions.\\

\subsection{Evaluation of intermolecular interaction energies}

\begin{figure}[ht!]
 \begin{center}
 \resizebox{0.75\columnwidth}{!}{\includegraphics[scale=0.5,clip,angle=0]{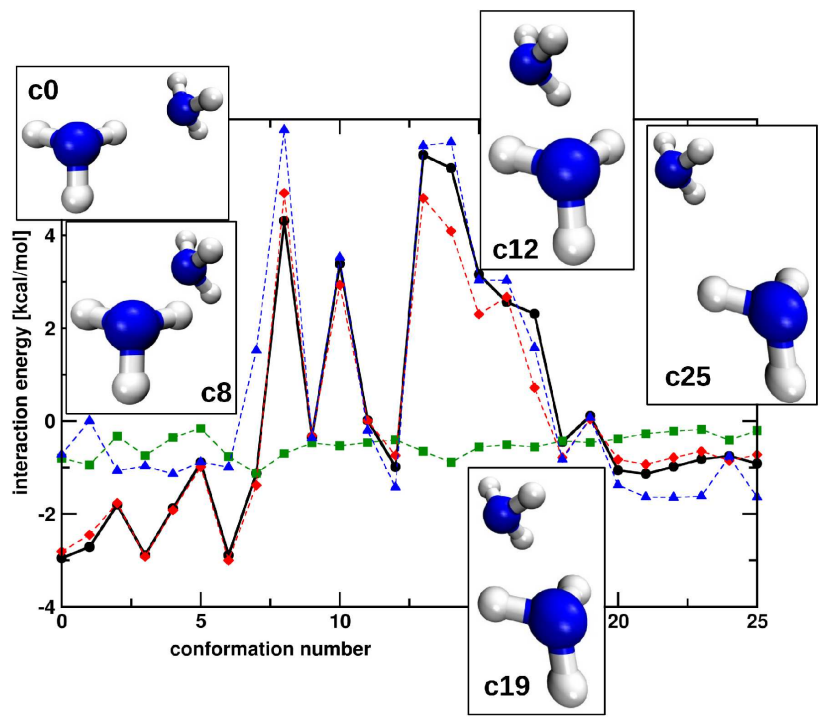}}
 \caption{\small NH$_3$ interaction energies fitted and compared to MP2/aug-cc-pVTZ 
 calculations (black): Model 1 (green squares), Model 2 (red circles), Model 3 (blue triangles).  Example conformations are shown with the corresponding conformation number.}
\label{fig:NH3-interactions}
\end{center}
\end{figure}

\begin{figure}[ht!]
 \begin{center}
 \resizebox{0.75\columnwidth}{!}{\includegraphics[scale=0.5,clip,angle=0]{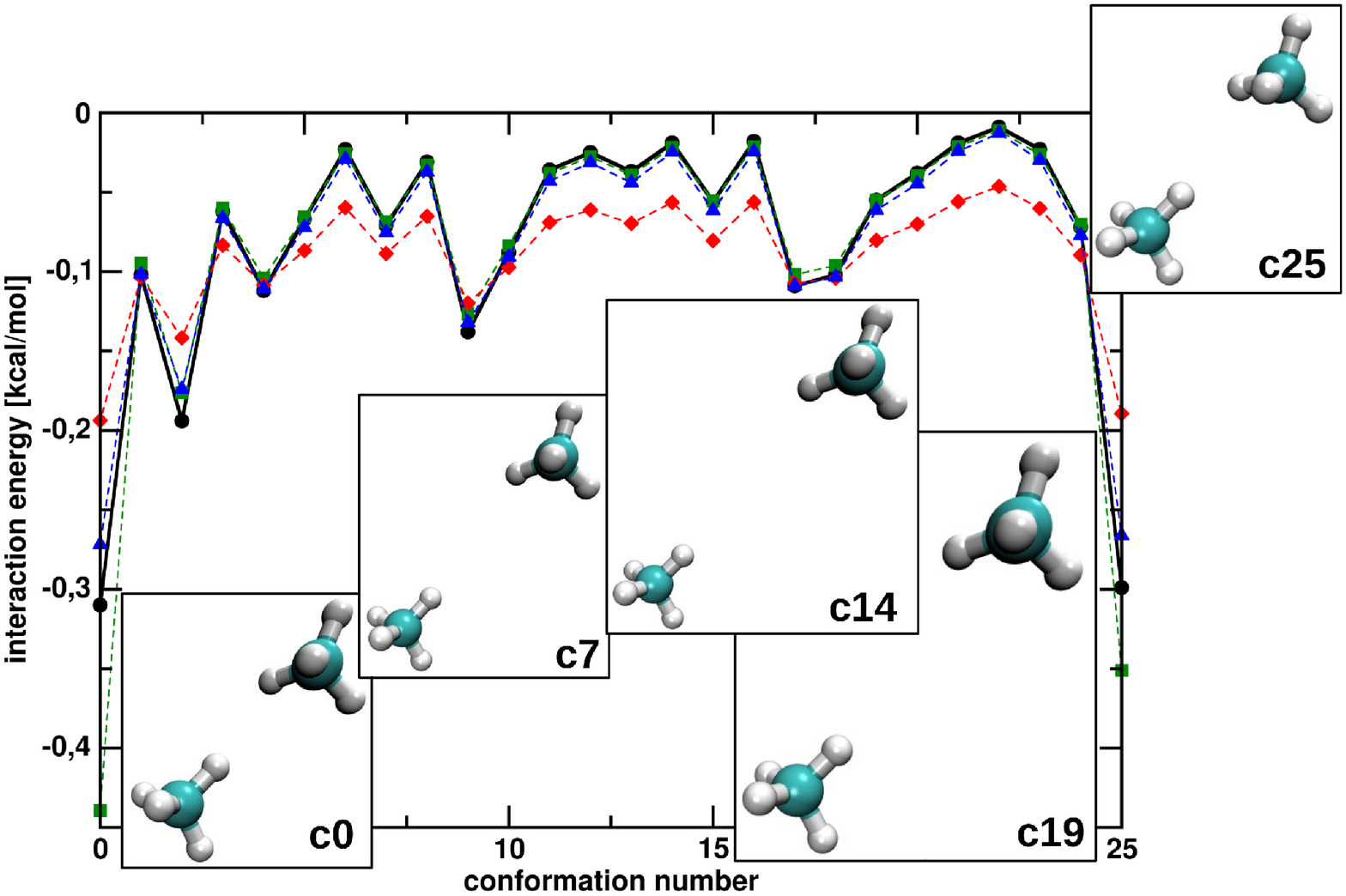}}
 \caption{\small CH$_4$ interaction energies fitted and compared to MP2/aug-cc-pVTZ 
 calculations (black): Model 1 (green squares), Model 2 (red circles), Model 3 (blue triangles).  Example conformations are shown with the corresponding conformation number.}
\label{fig:CH4-interactions}
\end{center}
\end{figure}

\begin{figure}[ht!]
 \begin{center}
 \resizebox{0.75\columnwidth}{!}{\includegraphics[scale=0.5,clip,angle=0]{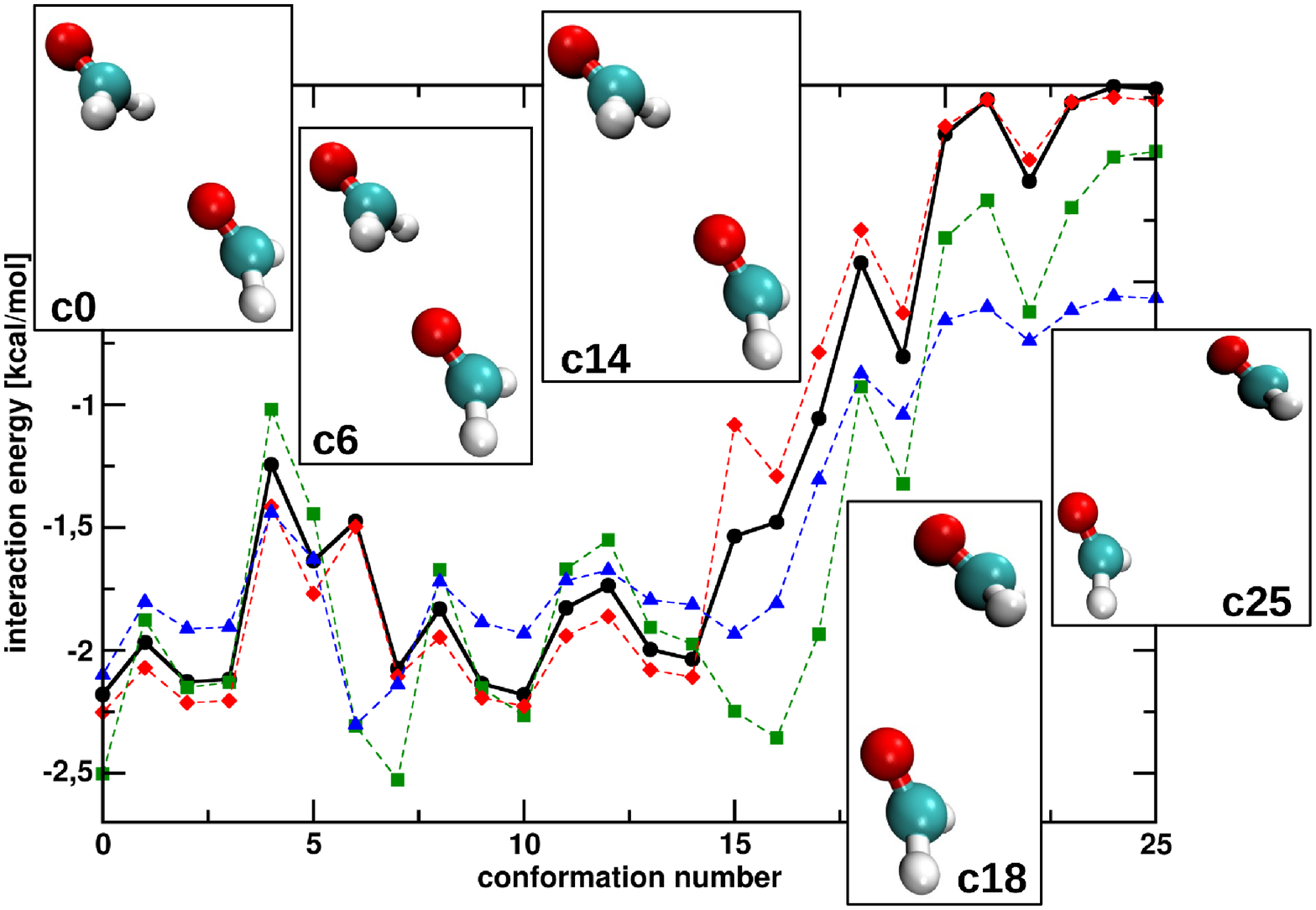}}
 \caption{\small Formaldehyde (OCH$_2$) interaction energies fitted and compared to MP2/aug-cc-pVTZ 
 calculations (black): Model 1 (green squares), Model 2 (red circles), Model 3 (blue triangles).  Example conformations are shown with the corresponding conformation number.}
\label{fig:OCH2-interactions}
\end{center}
\end{figure}

\noindent The electrostatic potential evaluations are suitable to assess the general properties of the different models. For the use of these models in atomistic force fields it is however more important to evaluate the accuracy of the intermolecular interaction energies. In principle 
the parameters obtained for the EPS could also be used for the interaction energies, however, in practice it has been found that it is better to parameterize force fields based on interaction energies and to jointly fit all parameters to the intermolecular energy.\cite{KramerMeuwlyJCTC2013} Therefore the $\zeta$- and $\epsilon$-parameters have 
been fitted to the intermolecular interaction energies of eight example molecules (Table \ref{tb:params2}), including both, molecules with only two atom types and molecules with more than two atom types. The accuracy of the three models compared to 
{\it ab initio} is first illustrated for NH$_3$, CH$_4$ and formaldehyde as example molecules. The interaction energies and a subset of the dimer geometries are shown in Figures \ref{fig:NH3-interactions}, 
\ref{fig:CH4-interactions} and \ref{fig:OCH2-interactions}.\\

\noindent Comparison of the three molecules shows a clearly visible advantage for Model 2 in the case of NH$_3$ and 
formaldehyde, whereas for methane Model 3 is the most accurate. The average energy differences $\bar{\Delta}_{E\_inter}$ 
between the {\it ab initio} interaction energies for all eight molecules are shown in Table \ref{tb:e-diff}. For all 
cases the most accurate model is either Model 2 or Model 3, i.e.one of the two models containing electron potential sites. This demonstrates the advantage of using electron potentials despite the fact that a number of generic parameters are
used and the number of parameters to be fitted is the same as 
for Model 1. Model 2 seems to perform better for polar 
molecules, i.e. for molecules with larger electronegativity differences, whereas Model 3 performs better for apolar molecules. A comparison of the $\zeta$-parameters obtained from the EPS in Table \ref{tb:params1} to the parameters obtained from the intermolecular interaction energies in Table \ref{tb:params2} shows that there is no systematic 
difference between the two sets of parameters for Models 1 and 2, whereas for Model 3 the parameters obtained from the interaction energies are systematically smaller. This indicates that Model 3 captures a systematic difference between the electrostatic potential of monomers and dimers, most likely due to polarization. Potentially this result could be interesting for describing polarization effects systematically. For Model 1 and 2, the differences in the parameters are not systematic and therefore most likely arising due to a combination of polarization effects and error compensation.

\begin{table}[h] 
\caption{Average energy differences $\bar{\Delta}_{E\_inter}$ between the {\it ab initio} interaction energies calculated at the MP2/aug-cc-pVTZ level of theory and each model in its point charge version.}
\begin{center}
\begin{tabular}{l||c|c|c} 
    &\multicolumn{3}{c}{$\bar{\Delta}_{E\_inter}$ kcal/mol} \\ 
\hline
    & Model 1 & Model 2 & Model 3 \\
\hline   
\hline
H$_2$O & 1.0198 & 0.9897 & 1.3504 \\
\hline
NH$_3$ & 1.8860 & 0.3326 & 0.8152 \\
\hline
CH$_4$ & 0.0107 & 0.0320 & 0.0078 \\
\hline
H$_2$S & 1.5788 & 0.6627 & 0.6166 \\
\hline
PH$_3$ & 0.0141 & 0.0178 & 0.0116  \\
\hline
CO$_2$ & 0.3651 & 0.2853 & 0.3317 \\
\hline
CH$_5$N & 0.0047 & 0.0046 & 0.0254 \\
\hline
OCH$_2$ & 0.3358 & 0.1060 & 0.3631  \\
\hline
\end{tabular}	            
\end{center}
\label{tb:e-diff}
\end{table}

\clearpage

\section{Discussion and Conclusions}

\noindent The possibility of constructing valence electron based potentials for the nonbonded interactions in atomistic force fields has been explored in this paper. Three charge distribution models using simple potential functions and only one adjustable parameter for the electrostatic potential of a molecule have been introduced and compared. It was shown that even with this simple parametrization which uses empirical constants and generic parameter values for the additional sites, the electrostatic potential is more or equally accurate than with population derived charges for all three models. For half of the evaluated molecules the EPS is equally or more accurate than distributed multipole moments for at least one of the models. The comparison of point charge interaction potentials to spherical Gaussian distributions showed that the accuracy of the electrostatic potential can be further increased by using Gaussian distributions. However, in most cases the improvement due to Gaussian functions is small and only relevant in the distance range close to the atoms. Since in this distance range the interactions are dominated by Pauli repulsion it is more important to assess the overall accuracy of the intermolecular interactions composed of electrostatics, 
LJ-potentials and repulsive potentials on the electron potential sites. The evaluation of the intermolecular interaction energies compared to {\it ab initio} revealed a systematic advantage of having electron potential sites. For the charge distribution models it was found that Model 2 which uses a polar charge distribution scheme is more accurate for molecules with larger electronegativity differences i.e. polar molecules, whereas Model 3 is more accurate for molecule with small electronegativity differences.\\

\noindent As all the molecules used for the evaluations here were small, it was possible to use molecular parameters for fitting the different molecular parameters. In order to generalize this approach to larger molecules it would most likely be necessary to use fitting parameters for groups of atoms rather than entire molecules. As the accuracy of 
the different models seems to depend on the polarity of the molecules, it would probably be suitable to parametrize larger molecules by partitioning them into functional groups. Overall the methods presented here offers a new concept for introducing and parametrizing additional interaction sites to improve the accuracy of intermolecular interactions in atomistic force fields. The concept of charge distribution models provides a pathway to use more 
accurate potential functions without increasing the parameter space and therefore the parametrization effort.\\

\section*{Supplementary Material}
\noindent See supplemetary material for the evaluations of the accuracy of the electrostatic potential compared for each model separately.\\

\section*{Acknowledgments}
\noindent The author wishes to acknowledge Dr. Mike Kuiper (University of Melbourne) for questioning the building principles of commonly used empirical force fields. Financial support from the Einstein Foundation Berlin through postdoctoral fellowship SOoPiC is gratefully acknowledged.\\

\bibliography{refs}

\begin{thebibliography}{35}
\expandafter\ifx\csname natexlab\endcsname\relax\def\natexlab#1{#1}\fi
\expandafter\ifx\csname bibnamefont\endcsname\relax
  \def\bibnamefont#1{#1}\fi
\expandafter\ifx\csname bibfnamefont\endcsname\relax
  \def\bibfnamefont#1{#1}\fi
\expandafter\ifx\csname citenamefont\endcsname\relax
  \def\citenamefont#1{#1}\fi
\expandafter\ifx\csname url\endcsname\relax
  \def\url#1{\texttt{#1}}\fi
\expandafter\ifx\csname urlprefix\endcsname\relax\def\urlprefix{URL }\fi
\providecommand{\bibinfo}[2]{#2}
\providecommand{\eprint}[2][]{\url{#2}}

\bibitem[{\citenamefont{{MacKerell, Jr.} et~al.}(1998)\citenamefont{{MacKerell,
  Jr.}, Bashford, Bellott, {Dunbrack, Jr.}, Evanseck, Field, Fischer, Gao, Guo,
  Ha et~al.}}]{charmm22}
\bibinfo{author}{\bibfnamefont{A.~D.} \bibnamefont{{MacKerell, Jr.}}},
  \bibinfo{author}{\bibfnamefont{D.}~\bibnamefont{Bashford}},
  \bibinfo{author}{\bibfnamefont{M.}~\bibnamefont{Bellott}},
  \bibinfo{author}{\bibfnamefont{R.~L.} \bibnamefont{{Dunbrack, Jr.}}},
  \bibinfo{author}{\bibfnamefont{J.~D.} \bibnamefont{Evanseck}},
  \bibinfo{author}{\bibfnamefont{M.~J.} \bibnamefont{Field}},
  \bibinfo{author}{\bibfnamefont{S.}~\bibnamefont{Fischer}},
  \bibinfo{author}{\bibfnamefont{J.}~\bibnamefont{Gao}},
  \bibinfo{author}{\bibfnamefont{H.}~\bibnamefont{Guo}},
  \bibinfo{author}{\bibfnamefont{S.}~\bibnamefont{Ha}}, \bibnamefont{et~al.},
  \bibinfo{journal}{J. Phys. Chem. B} \textbf{\bibinfo{volume}{102}},
  \bibinfo{pages}{3586} (\bibinfo{year}{1998}).

\bibitem[{\citenamefont{Weiner et~al.}(1984)\citenamefont{Weiner, Kollman,
  Case, Singh, Ghio, Alagona, {Profeta Jr}, and Weiner}}]{amber}
\bibinfo{author}{\bibfnamefont{S.~J.} \bibnamefont{Weiner}},
  \bibinfo{author}{\bibfnamefont{P.~A.} \bibnamefont{Kollman}},
  \bibinfo{author}{\bibfnamefont{D.~A.} \bibnamefont{Case}},
  \bibinfo{author}{\bibfnamefont{U.}~\bibnamefont{Singh}},
  \bibinfo{author}{\bibfnamefont{C.}~\bibnamefont{Ghio}},
  \bibinfo{author}{\bibfnamefont{G.}~\bibnamefont{Alagona}},
  \bibinfo{author}{\bibfnamefont{S.}~\bibnamefont{{Profeta Jr}}},
  \bibnamefont{and} \bibinfo{author}{\bibfnamefont{P.}~\bibnamefont{Weiner}},
  \bibinfo{journal}{J. Am. Chem. Soc.} \textbf{\bibinfo{volume}{106}},
  \bibinfo{pages}{765} (\bibinfo{year}{1984}).

\bibitem[{\citenamefont{Van~Gunstern and Berendsen}(1987)}]{GROMOS}
\bibinfo{author}{\bibfnamefont{W.~F.} \bibnamefont{Van~Gunstern}}
  \bibnamefont{and} \bibinfo{author}{\bibfnamefont{H.~J.~C.}
  \bibnamefont{Berendsen}} (\bibinfo{year}{1987}).

\bibitem[{\citenamefont{Jorgensen and Tirado-Rives}(1988)}]{OPLS}
\bibinfo{author}{\bibfnamefont{W.~L.} \bibnamefont{Jorgensen}}
  \bibnamefont{and}
  \bibinfo{author}{\bibfnamefont{J.}~\bibnamefont{Tirado-Rives}},
  \bibinfo{journal}{J. Am. Chem. Soc.} \textbf{\bibinfo{volume}{110}},
  \bibinfo{pages}{1657} (\bibinfo{year}{1988}).

\bibitem[{\citenamefont{Karplus and Petsko}(1990)}]{Karplus90}
\bibinfo{author}{\bibfnamefont{M.}~\bibnamefont{Karplus}} \bibnamefont{and}
  \bibinfo{author}{\bibfnamefont{G.~A.} \bibnamefont{Petsko}},
  \bibinfo{journal}{Nature} \textbf{\bibinfo{volume}{347}},
  \bibinfo{pages}{631} (\bibinfo{year}{1990}).

\bibitem[{\citenamefont{Van~Gunstern et~al.}(1994)\citenamefont{Van~Gunstern,
  Luque, Timms, and Torda}}]{vanGunstern94}
\bibinfo{author}{\bibfnamefont{W.~F.} \bibnamefont{Van~Gunstern}},
  \bibinfo{author}{\bibfnamefont{F.~J.} \bibnamefont{Luque}},
  \bibinfo{author}{\bibfnamefont{D.}~\bibnamefont{Timms}}, \bibnamefont{and}
  \bibinfo{author}{\bibfnamefont{A.~E.} \bibnamefont{Torda}},
  \bibinfo{journal}{Ann. Rev. Biophys. Biomol. Struct.}
  \textbf{\bibinfo{volume}{23}}, \bibinfo{pages}{847} (\bibinfo{year}{1994}).

\bibitem[{\citenamefont{Price and Stone}(1992)}]{Price92}
\bibinfo{author}{\bibfnamefont{S.}~\bibnamefont{Price}} \bibnamefont{and}
  \bibinfo{author}{\bibfnamefont{A.}~\bibnamefont{Stone}}, \bibinfo{journal}{J.
  Chem. Soc. Faraday Trans.} \textbf{\bibinfo{volume}{88}},
  \bibinfo{pages}{1755} (\bibinfo{year}{1992}).

\bibitem[{\citenamefont{Hodges et~al.}(1997)\citenamefont{Hodges, Stone, and
  Xantheas}}]{HodgesStone97}
\bibinfo{author}{\bibfnamefont{M.~P.} \bibnamefont{Hodges}},
  \bibinfo{author}{\bibfnamefont{A.~J.} \bibnamefont{Stone}}, \bibnamefont{and}
  \bibinfo{author}{\bibfnamefont{S.~S.} \bibnamefont{Xantheas}},
  \bibinfo{journal}{J. Phys. Chem. A} \textbf{\bibinfo{volume}{101}},
  \bibinfo{pages}{9163} (\bibinfo{year}{1997}).

\bibitem[{\citenamefont{Kedzierski and Sokalski}(2001)}]{Sokalski01}
\bibinfo{author}{\bibfnamefont{P.}~\bibnamefont{Kedzierski}} \bibnamefont{and}
  \bibinfo{author}{\bibfnamefont{W.}~\bibnamefont{Sokalski}},
  \bibinfo{journal}{J. Comp. Chem.} \textbf{\bibinfo{volume}{22}},
  \bibinfo{pages}{1082} (\bibinfo{year}{2001}).

\bibitem[{\citenamefont{Karamertzanis and Price}(2006)}]{Price06}
\bibinfo{author}{\bibfnamefont{P.~G.} \bibnamefont{Karamertzanis}}
  \bibnamefont{and} \bibinfo{author}{\bibfnamefont{S.~L.} \bibnamefont{Price}},
  \bibinfo{journal}{J. Chem. Theory Comput.} \textbf{\bibinfo{volume}{2}},
  \bibinfo{pages}{1184} (\bibinfo{year}{2006}).

\bibitem[{\citenamefont{Plattner and Meuwly}(2009)}]{JMolM09}
\bibinfo{author}{\bibfnamefont{N.}~\bibnamefont{Plattner}} \bibnamefont{and}
  \bibinfo{author}{\bibfnamefont{M.}~\bibnamefont{Meuwly}},
  \bibinfo{journal}{J. Mol. Model.} \textbf{\bibinfo{volume}{15}},
  \bibinfo{pages}{687} (\bibinfo{year}{2009}).

\bibitem[{\citenamefont{Kramer et~al.}(2013{\natexlab{a}})\citenamefont{Kramer,
  Bereau, Spinn, Liedl, Gedeck, and Meuwly}}]{KramerMeuwly2013}
\bibinfo{author}{\bibfnamefont{C.}~\bibnamefont{Kramer}},
  \bibinfo{author}{\bibfnamefont{T.}~\bibnamefont{Bereau}},
  \bibinfo{author}{\bibfnamefont{A.}~\bibnamefont{Spinn}},
  \bibinfo{author}{\bibfnamefont{K.~R.} \bibnamefont{Liedl}},
  \bibinfo{author}{\bibfnamefont{P.}~\bibnamefont{Gedeck}}, \bibnamefont{and}
  \bibinfo{author}{\bibfnamefont{M.}~\bibnamefont{Meuwly}},
  \bibinfo{journal}{J. Chem. Inf. Model.} \textbf{\bibinfo{volume}{53}},
  \bibinfo{pages}{3410} (\bibinfo{year}{2013}{\natexlab{a}}).

\bibitem[{\citenamefont{Cardamone et~al.}(2014)\citenamefont{Cardamone, Hughes,
  and Popelier}}]{Popelier2014}
\bibinfo{author}{\bibfnamefont{S.}~\bibnamefont{Cardamone}},
  \bibinfo{author}{\bibfnamefont{T.~J.} \bibnamefont{Hughes}},
  \bibnamefont{and} \bibinfo{author}{\bibfnamefont{P.~L.~A.}
  \bibnamefont{Popelier}}, \bibinfo{journal}{Phys. Chem. Chem. Phys.}
  \textbf{\bibinfo{volume}{16}}, \bibinfo{pages}{10367} (\bibinfo{year}{2014}).

\bibitem[{\citenamefont{Mahoney and Jorgensen}(2000)}]{TIP5P}
\bibinfo{author}{\bibfnamefont{M.~W.} \bibnamefont{Mahoney}} \bibnamefont{and}
  \bibinfo{author}{\bibfnamefont{W.~L.} \bibnamefont{Jorgensen}},
  \bibinfo{journal}{J. Chem. Phys.} \textbf{\bibinfo{volume}{112}},
  \bibinfo{pages}{8910} (\bibinfo{year}{2000}).

\bibitem[{\citenamefont{Straub and Karplus}(1991)}]{StraubKarplus1991}
\bibinfo{author}{\bibfnamefont{J.~E.} \bibnamefont{Straub}} \bibnamefont{and}
  \bibinfo{author}{\bibfnamefont{M.}~\bibnamefont{Karplus}},
  \bibinfo{journal}{Chem. Phys.} \textbf{\bibinfo{volume}{158}},
  \bibinfo{pages}{221} (\bibinfo{year}{1991}).

\bibitem[{\citenamefont{Saxena and Sept}(2013)}]{SaxenaSept2013}
\bibinfo{author}{\bibfnamefont{A.}~\bibnamefont{Saxena}} \bibnamefont{and}
  \bibinfo{author}{\bibfnamefont{D.}~\bibnamefont{Sept}}, \bibinfo{journal}{J.
  Chem. Theory Comput.} \textbf{\bibinfo{volume}{9}}, \bibinfo{pages}{3538}
  (\bibinfo{year}{2013}).

\bibitem[{\citenamefont{Lopes et~al.}(2013)\citenamefont{Lopes, Huang, Shim,
  Luo, Li, Roux, and {MacKerell, Jr.}}}]{LopesMacKerell2013}
\bibinfo{author}{\bibfnamefont{P.~E.~M.} \bibnamefont{Lopes}},
  \bibinfo{author}{\bibfnamefont{J.}~\bibnamefont{Huang}},
  \bibinfo{author}{\bibfnamefont{J.}~\bibnamefont{Shim}},
  \bibinfo{author}{\bibfnamefont{Y.}~\bibnamefont{Luo}},
  \bibinfo{author}{\bibfnamefont{H.}~\bibnamefont{Li}},
  \bibinfo{author}{\bibfnamefont{B.}~\bibnamefont{Roux}}, \bibnamefont{and}
  \bibinfo{author}{\bibfnamefont{A.~D.} \bibnamefont{{MacKerell, Jr.}}},
  \bibinfo{journal}{J. Chem. Theory Comput.} \textbf{\bibinfo{volume}{9}},
  \bibinfo{pages}{5430} (\bibinfo{year}{2013}).

\bibitem[{\citenamefont{Ren and Ponder}(2003)}]{AMOEBA}
\bibinfo{author}{\bibfnamefont{P.}~\bibnamefont{Ren}} \bibnamefont{and}
  \bibinfo{author}{\bibfnamefont{J.~W.} \bibnamefont{Ponder}},
  \bibinfo{journal}{J. Phys. Chem. B} \textbf{\bibinfo{volume}{107}},
  \bibinfo{pages}{5933} (\bibinfo{year}{2003}).

\bibitem[{\citenamefont{Gresh et~al.}(2007)\citenamefont{Gresh, Cisneros, and
  Darden}}]{SIBFA}
\bibinfo{author}{\bibfnamefont{N.}~\bibnamefont{Gresh}},
  \bibinfo{author}{\bibfnamefont{A.~G.} \bibnamefont{Cisneros}},
  \bibnamefont{and} \bibinfo{author}{\bibfnamefont{J.-P.} \bibnamefont{Darden},
  \bibfnamefont{T.~A.~Piquemal}}, \bibinfo{journal}{J. Chem. Theory Comput.}
  \textbf{\bibinfo{volume}{3}}, \bibinfo{pages}{1960} (\bibinfo{year}{2007}).

\bibitem[{\citenamefont{Devereux et~al.}(2014)\citenamefont{Devereux,
  Raghunathan, Fedorov, and Meuwly}}]{DevereuxMeuwly2014}
\bibinfo{author}{\bibfnamefont{M.}~\bibnamefont{Devereux}},
  \bibinfo{author}{\bibfnamefont{S.}~\bibnamefont{Raghunathan}},
  \bibinfo{author}{\bibfnamefont{D.~G.} \bibnamefont{Fedorov}},
  \bibnamefont{and} \bibinfo{author}{\bibfnamefont{M.}~\bibnamefont{Meuwly}},
  \bibinfo{journal}{J. Chem. Theory Comput.} \textbf{\bibinfo{volume}{10}},
  \bibinfo{pages}{4229} (\bibinfo{year}{2014}).

\bibitem[{\citenamefont{Stone}(1996)}]{Stone96}
\bibinfo{author}{\bibfnamefont{A.~J.} \bibnamefont{Stone}},
  \emph{\bibinfo{title}{The Theory of Intermolecular Forces}}
  (\bibinfo{publisher}{Clarendon Press}, \bibinfo{address}{Oxford},
  \bibinfo{year}{1996}).

\bibitem[{\citenamefont{Su and {Goddard III}}(2007)}]{SuGoddard2007}
\bibinfo{author}{\bibfnamefont{J.~T.} \bibnamefont{Su}} \bibnamefont{and}
  \bibinfo{author}{\bibfnamefont{W.~A.} \bibnamefont{{Goddard III}}},
  \bibinfo{journal}{Phys. Rev. Lett.} \textbf{\bibinfo{volume}{99}},
  \bibinfo{pages}{185003} (\bibinfo{year}{2007}).

\bibitem[{\citenamefont{Jaramillo-Botero
  et~al.}(2011)\citenamefont{Jaramillo-Botero, Su, and
  Qi}}]{Jaramillo-BoteroGoddard2011}
\bibinfo{author}{\bibfnamefont{A.}~\bibnamefont{Jaramillo-Botero}},
  \bibinfo{author}{\bibfnamefont{J.~T.} \bibnamefont{Su}}, \bibnamefont{and}
  \bibinfo{author}{\bibfnamefont{W.~A.} \bibnamefont{Qi},
  \bibfnamefont{A.~{Goddard III}}}, \bibinfo{journal}{J. Comp. Chem.}
  \textbf{\bibinfo{volume}{32}}, \bibinfo{pages}{497} (\bibinfo{year}{2011}).

\bibitem[{\citenamefont{Slater}(1964)}]{Slater64}
\bibinfo{author}{\bibfnamefont{J.~C.} \bibnamefont{Slater}},
  \bibinfo{journal}{J. Chem. Phys.} \textbf{\bibinfo{volume}{41}},
  \bibinfo{pages}{3199} (\bibinfo{year}{1964}).

\bibitem[{\citenamefont{Kramer et~al.}(2013{\natexlab{b}})\citenamefont{Kramer,
  Gedeck, and Meuwly}}]{KramerMeuwlyJCTC2013}
\bibinfo{author}{\bibfnamefont{C.}~\bibnamefont{Kramer}},
  \bibinfo{author}{\bibfnamefont{P.}~\bibnamefont{Gedeck}}, \bibnamefont{and}
  \bibinfo{author}{\bibfnamefont{M.}~\bibnamefont{Meuwly}},
  \bibinfo{journal}{J. Chem. Theory Comput.} \textbf{\bibinfo{volume}{9}},
  \bibinfo{pages}{1499} (\bibinfo{year}{2013}{\natexlab{b}}).

\bibitem[{\citenamefont{Pauling}(1932)}]{Pauling32}
\bibinfo{author}{\bibfnamefont{L.}~\bibnamefont{Pauling}}, \bibinfo{journal}{J.
  Am. Chem. Soc.} \textbf{\bibinfo{volume}{54}}, \bibinfo{pages}{3570}
  (\bibinfo{year}{1932}).

\bibitem[{\citenamefont{Bondi}(1964)}]{Bondi64}
\bibinfo{author}{\bibfnamefont{A.}~\bibnamefont{Bondi}}, \bibinfo{journal}{J.
  Phys. Chem.} \textbf{\bibinfo{volume}{68}}, \bibinfo{pages}{441}
  (\bibinfo{year}{1964}).

\bibitem[{\citenamefont{Mulliken}(1934{\natexlab{a}})}]{Mulliken34}
\bibinfo{author}{\bibfnamefont{R.}~\bibnamefont{Mulliken}},
  \bibinfo{journal}{J. Chem. Phys.} \textbf{\bibinfo{volume}{2}},
  \bibinfo{pages}{782} (\bibinfo{year}{1934}{\natexlab{a}}).

\bibitem[{\citenamefont{Neese}(2012)}]{ORCA}
\bibinfo{author}{\bibfnamefont{F.}~\bibnamefont{Neese}},
  \bibinfo{journal}{Wiley Interdiscip. Rev.: Comput. Mol. Sci.}
  \textbf{\bibinfo{volume}{2}}, \bibinfo{pages}{73} (\bibinfo{year}{2012}).

\bibitem[{\citenamefont{Becke}(1993)}]{Becke93}
\bibinfo{author}{\bibfnamefont{A.~D.} \bibnamefont{Becke}},
  \bibinfo{journal}{J. Chem. Phys.} \textbf{\bibinfo{volume}{98}},
  \bibinfo{pages}{5648} (\bibinfo{year}{1993}).

\bibitem[{\citenamefont{Dunning}(1989)}]{Dun89}
\bibinfo{author}{\bibfnamefont{T.~H.} \bibnamefont{Dunning},
  \bibfnamefont{Jr.}}, \bibinfo{journal}{J. Chem. Phys.}
  \textbf{\bibinfo{volume}{90}}, \bibinfo{pages}{1007} (\bibinfo{year}{1989}).

\bibitem[{\citenamefont{Mulliken}(1934{\natexlab{b}})}]{Mulliken55}
\bibinfo{author}{\bibfnamefont{R.}~\bibnamefont{Mulliken}},
  \bibinfo{journal}{J. Chem. Phys.} \textbf{\bibinfo{volume}{23}},
  \bibinfo{pages}{1833} (\bibinfo{year}{1934}{\natexlab{b}}).

\bibitem[{\citenamefont{Frisch et~al.}(2004)\citenamefont{Frisch, Trucks,
  Schlegel, Scuseria, Robb, Cheeseman, Montgomery, Jr., Vreven, Kudin
  et~al.}}]{Gaussian03}
\bibinfo{author}{\bibfnamefont{M.~J.} \bibnamefont{Frisch}},
  \bibinfo{author}{\bibfnamefont{G.~W.} \bibnamefont{Trucks}},
  \bibinfo{author}{\bibfnamefont{H.~B.} \bibnamefont{Schlegel}},
  \bibinfo{author}{\bibfnamefont{G.~E.} \bibnamefont{Scuseria}},
  \bibinfo{author}{\bibfnamefont{M.~A.} \bibnamefont{Robb}},
  \bibinfo{author}{\bibfnamefont{J.~R.} \bibnamefont{Cheeseman}},
  \bibinfo{author}{\bibfnamefont{J.~A.} \bibnamefont{Montgomery}},
  \bibinfo{author}{\bibnamefont{Jr.}},
  \bibinfo{author}{\bibfnamefont{T.}~\bibnamefont{Vreven}},
  \bibinfo{author}{\bibfnamefont{K.~N.} \bibnamefont{Kudin}},
  \bibnamefont{et~al.} (\bibinfo{year}{2004}).

\bibitem[{\citenamefont{Stone}(2005)}]{GDMA05}
\bibinfo{author}{\bibfnamefont{A.~J.} \bibnamefont{Stone}},
  \bibinfo{journal}{J. Chem. Theory Comput.} \textbf{\bibinfo{volume}{1}},
  \bibinfo{pages}{1128} (\bibinfo{year}{2005}).

\bibitem[{\citenamefont{Simon et~al.}(1996)\citenamefont{Simon, Duran, and
  Dannenberg}}]{SimonDannenberg96}
\bibinfo{author}{\bibfnamefont{S.}~\bibnamefont{Simon}},
  \bibinfo{author}{\bibfnamefont{M.}~\bibnamefont{Duran}}, \bibnamefont{and}
  \bibinfo{author}{\bibfnamefont{J.~J.} \bibnamefont{Dannenberg}},
  \bibinfo{journal}{J. Chem. Phys.} \textbf{\bibinfo{volume}{105}},
  \bibinfo{pages}{11024} (\bibinfo{year}{1996}).

\end{thebibliography}
\clearpage

\end{document}